\documentclass[prd,preprintnumbers,12pt]{revtex4}
\usepackage{epsfig}
\usepackage{graphicx}

\newcommand{\be}{\begin{equation}}

\newcommand{\ee}{\end{equation}}
\newcommand{\bea}{\begin{eqnarray}}
\newcommand{\eea}{\end{eqnarray}}

\newcommand{\nn}{\nonumber}
\newcommand{\de}{\partial}

\begin{document}


\preprint{{\bf CERN-PH-TH-07-160}} \preprint{{\bf UB-ECM-PF-07-20}}
\title{Exotic Non-relativistic String}

\author{Roberto Casalbuoni}
\email[Electronic address: ]{casalbuoni@fi.infn.it}
\affiliation{Department of Physics, University of Florence}
\affiliation{INFN, Florence, Italy} \affiliation{Galileo
Galilei Institute for Theoretical Physics, Florence, Italy}
\author{Joaquim Gomis}
\email[Electronic address: ]{gomis@ecm.ub.es}
 \affiliation{PH-TH Division
CERN\\CH-1211 Geneva 23, Switzerland}
\affiliation{Departament d'Estructura i Constituents de la Mat\`eria,\\
Facultat de F\'isica, Universitat de Barcelona,\\
Diagonal 647, 08028 Barcelona, Spain}
\author{Giorgio Longhi}
\email[Electronic address: ]{longhi@fi.infn.it}
\affiliation{Department of Physics, University of Florence}
\affiliation{INFN, Florence, Italy}

\begin{abstract}
We construct a classical non-relativistic string model in
3+1 dimensions. The model contains a spurion tensor field
that is responsible for the non-commutative structure of
the model. Under double dimensional reduction the model
reduces to the exotic non-relativistic particle in 2+1
dimensions.

\end{abstract}

\newcommand{\f}[2]{\frac{#1}{#2}}
\def\to{\rightarrow}
\def\ptl{\partial}
\def\beq{\begin{equation}}
\def\eeq{\end{equation}}
\def\bea{\begin{eqnarray}}
\def\eea{\end{eqnarray}}
\def\nn{\nonumber}
\def\half{{1\over 2}}
\def\rhalf{{1\over \sqrt 2}}
\def\calo{{\cal O}}
\def\call{{\cal L}}
\def\calm{{\cal M}}
\def\del{\delta}
\def\eps{\epsilon}
\def\lam{\lambda}

\def\anti{\overline}
\def\delfac{\sqrt{{2(\del-1)\over 3(\del+2)}}}
\def\heff{h'}
\def\square{\boxxit{0.4pt}{\fillboxx{7pt}{7pt}}\hspace*{1pt}}
    \def\boxxit#1#2{\vbox{\hrule height #1 \hbox {\vrule width #1
             \vbox{#2}\vrule width #1 }\hrule height #1 } }
    \def\fillboxx#1#2{\hbox to #1{\vbox to #2{\vfil}\hfil}   }

\def\braket#1#2{\langle #1| #2\rangle}
\def\gev{~{\rm GeV}}
\def\gam{\gamma}
\def\sn{s_{\vec n}}
\def\sm{s_{\vec m}}
\def\mm{m_{\vec m}}
\def\mn{m_{\vec n}}
\def\mh{m_h}
\def\sumn{\sum_{\vec n>0}}
\def\summ{\sum_{\vec m>0}}
\def\vl{\vec l}
\def\vk{\vec k}
\def\ml{m_{\vl}}
\def\mk{m_{\vk}}
\def\gp{g'}
\def\gt{\tilde g}
\def\hw{{\hat W}}
\def\hz{{\hat Z}}
\def\ha{{\hat A}}

\def\yy{{\cal Y}_\mu}
\def\yyt{{\tilde{\cal Y}}_\mu}
\def\lq{\left [}
\def\rq{\right ]}
\def\dmu{\partial_\mu}
\def\dnu{\partial_\nu}
\def\dmus{\partial^\mu}
\def\dnus{\partial^\nu}
\def\gp{g'}
\def\gpt{{\tilde g'}}
\def\gs{g''}
\def\ggs{\frac{g}{\gs}}
\def\eps{{\epsilon}}
\def\tr{{\rm {tr}}}
\def\V{{\bf{V}}}
\def\W{{\bf{W}}}
\def\Wt{\tilde{ {W}}}
\def\Y{{\bf{Y}}}
\def\Yt{\tilde{ {Y}}}
\def\L{{\cal L}}
\def\s{\sigma}
\def\st{s_{\tilde\theta}}
\def\c{c_\theta}
\def\ct{c_{\tilde\theta}}
\def\gt{\tilde g}
\def\et{\tilde e}
\def\At{\tilde A}
\def\Zt{\tilde Z}
\def\Wpt{\tilde W^+}
\def\Wmt{\tilde W^-}

\maketitle

\section{Introduction \label{sec:0}}

The free massive non-relativistic particle in D dimensions
has as a symmetry group, the Galilei group, with a central
extension associated to the mass of the particle. Instead,
in 2+1 dimensions, there is a two-fold central extension
\cite{LL,BGO,BGGK}, where the second central element is
interpreted as a non-commutative parameter. These central
extensions are related to a nontrivial Eilenberg-Chevalley
cohomology of degree two. In fact every closed invariant
2-form that locally is the differential of a 1-form, which
is not invariant, produces an extension of the algebra, see
for example \cite{bargmann,levyleblond69,azcarragabook}.

 The exotic (2+1)D Galilei symmetry has appeared in the context
of non-commutative geometry and condensed matter physics
\cite{LSZ,DH1,JackN,DH2,HP1,PHMP,HP3,mariano1,OP,Horv,Polychronakos:2007df}.
It is a symmetry of a free relativistic particle in a
non-commutative plane in a special non-relativistic limit
\cite{JackN,HP1,PHMP}.

Instead, if one consider the  non-relativistic limit of a
relativistic particle in the AdS  in 2+1 dimensions, one
finds a system with an  exotic Newton-Hooke symmetry
\cite{Mariano}. The system has three essentially different
phases \cite{Alvarez:2007fw}, depending on the values of
the two central charges, which are present in the model.
The subcritical and supercritical phases (describing 2D
isotropic ordinary and exotic oscillators) are separated by
the critical phase (one-mode oscillator).

Non-relativistic extended objects  has been recently
considered in the literature. In particular,
non-relativistic string theory \cite{Gomis:2000bd} (see
also \cite{Danielsson:2000gi}) in flat space is  a
consistent decoupled sector of the bosonic string theory,
whose worldsheet, in the conformal field theory description
\cite{Gomis:2000bd}, possesses the appropriate Galilean
symmetry. Non-relativistic string theory can be derived as
a certain decoupling limit of the original relativistic
theory, even though the theory can be written down without
any reference to the original parent theory. The basic idea
behind the decoupling limit is to take a particular {\it
non-relativistic} limit, in such a way that only states
satisfying a Galilean invariant dispersion relation have
finite energy, while the rest of the states decouples.

The action of non-relativistic bosonic string theory, for a
flat D dimensional space-time, can be obtained with the
method of non-linear realizations \cite{Coleman} as a
Wess-Zumino (WS) term of the appropriate (string) Galilean
group \cite{Brugues:2004an,Brugues:2006yd}. The extended
algebra has non-central elements that transform
non-trivially under the stability group.

Particle models can be obtained by a double-dimensional
reduction of strings models. It is natural to ask if it is
possible to construct an exotic non-relativistic string in
3+1 dimensions, with other additional extended charges,
such that by reduction one could obtain the exotic particle
models appeared in the literature \cite{LSZ,DH1,HP1,PHMP}.
To this end we construct the classical action for exotic
Galilean non-relativistic strings, using the method of
non-linear realizations. We first construct the exotic
string Galilei algebra in 3+1 dimensions. This algebra
contains new extra extended generators. As we will see, the
parameters associated to these new extensions are related
to non-commutative parameters. The string action is
constructed from a 2-form, which is the product of two
invariant Maurer-Cartan (MC) 1-forms of the exotic extended
string Galilei algebra. The parameters of the string action
are the tension and a tensor parameter (spurion),
generalizing the magnetic field in the case of a particle.

We perform a classical Hamiltonian analysis of the model.
The hamiltonian turns out to be non-local in the original
variables. If we introduce an appropriate change of
variables, we can eliminate the non-locality, ending with a
theory formally equivalent to the ordinary non-relativistic
one for particular boundary conditions. Therefore, in this
last case, the symmetry group of the model in the new
variables is the non-exotic Galilei group.

The organization of the paper is as follows. In Section II
we consider the extended exotic Galilei algebra. In Section
III we introduce the string action and in Section IV we
study the dynamics in the static gauge, showing that there
are second class constraints in the model, then we proceed
to evaluate the Dirac brackets.  This analysis shows that
the Dirac brackets of the position variables are not zero,
giving rise to a non-commutative theory. In Section V we
show that it is possible, through a change of variables, to
make the model equivalent to the non relativistic string
model, only for particular choices of the boundary
conditions. We conclude with an outlook and a discussion in
Section VI.

\section{Exotic Extended Galilei String Algebra}

Let us consider a non-relativistic string in a D
dimensional flat space time.

We will denote by $x^0,x^1$ and $x^2,...x^{D-1} $ the
longitudinal (along the string) and transverse coordinates
respectively. The symmetry algebra of this system is
 the extended Galilei algebra, G
\cite{Brugues:2004an} \cite{Brugues:2006yd}, which is given
by

 \bea \left[P_\mu,P_\nu\right]&=&0\ , \qquad
\left[M_{\mu\nu},P_\rho\right]= i(\eta_{\mu\rho} P_{\nu} -
\eta_{\nu\rho} P_{\mu}) \ ,
\nonumber\\
\left[M_{\mu\nu},M_{\rho\s}\right]&=& 0 \ ,
\label{contra1}\\
\left[M_{ab},P_c\right]&=&i(\eta_{ac}P_b - \eta_{bc}P_a) \
,
\nonumber\\
\left[M_{ab},M_{cd}\right]&=&
i(\eta_{ac}M_{bd}+\eta_{bd}M_{ac}
 -\eta_{ad}M_{bc}-\eta_{bc}M_{ad}),
\label{contra3}\\
\left[P_\mu,P_a\right] &=&0\ , \qquad \left[M_{\mu
a},P_\nu\right] =i\eta_{\mu \nu}P_a \ ,
\nonumber\\
\left[M_{\mu\nu},M_{\rho c}\right]&=&i
(\eta_{\mu\rho}M_{\nu c}-\eta_{\nu\rho}M_{\mu c})\ ,\quad
\left[M_{\mu a},M_{bc}\right] = -i (\eta_{ab}M_{\mu
c}-\eta_{ac}M_{\mu b}), \label{contra2} \eea \bea \left[
P_{a}, ~ M_{\nu b}\right] &=& i\eta_{ab}\;Z_\nu, \qquad
\left[   M_{\mu a}, ~  M_{\nu b}\right]=i\eta_{ab}\;
\epsilon_{\mu\nu} Z, \label{contra4} \eea

and \bea \left[   P_\mu,~ Z_{\nu}\right]&=&0,\qquad \qquad
\nn\\
\left[   P_{\mu}, Z \right]&=&+ i\eta_{\mu\nu}
\epsilon^{\nu\lambda} Z_{\lambda},
\label{contra5}\\
 \left[  Z_{\mu},  M_{\rho\s} \right]&=&-i\eta_{\mu[\rho} Z_{\sigma]} \qquad
\left[ Z,  M_{\rho\s} \right]= 0. \label{contra6} \eea

In these expressions $P_\mu,\mu=0,1$, are the longitudinal
translation generators, $P_a, a=2,...D-1$ are the
transverse ones, $M_{\mu\nu}, M_{\mu a}, M_{a b}$ are the
generators of the boost transformation along the
longitudinal direction, generalized boosts and rotations in
the transverse space and $Z, Z_\mu$ are the extended
elements. Our conventions for the metric tensor are:
$\eta_{\mu\nu}={-,+},\eta_{ab}={+,+,...+}$. $[\mu \nu]$
means antisymmetry in the interchange of the two indices
$\mu \nu$. The Levi-Civita tensor is defined with
$\epsilon^{0 1} = + 1$.

Observe that the commutator of $M_{\mu\nu}$ and
$M_{\rho\sigma}$ is zero. This is due to the fact that in 2
dimensions ($\mu, \nu = 0, 1$) there is only one generator
of Lorentz transformations, and no rotations.

The action of non-relativistic string can be obtained by
the method of non-linear realizations
\cite{GomisKW,Brugues:2004an,Brugues:2006yd}. Let  us
consider the coset $G/{H}$, with the stability group $H$
generated by $(M_{\mu\nu},M_{a b})$.

Locally we parametrize a coset element as \footnote{The
introduction of a Goldstone field associated to a central
extension was first considered in \cite{Gauntlett:1990nk}}

\be \label{p3} g=e^{i x^{\mu}P_{\mu}}\;e^{i
x^{a}P_{a}}e^{iv^{\mu a}M_{\mu a}} \;e^{i c^\mu Z_\mu} e^{i
c Z}.\qquad \ee The (Goldstone) coordinates of the coset
depend on the parameters $\tau, \sigma$ of the world-sheet,

The MC 1-form is given by \be \Omega=-i g^{-1}dg= L^\mu
P_\mu+L^aP_a+L^{\mu a}M_{\mu a}+\frac 12 L^{\mu \nu} M_{\mu
\nu}+ \frac 12 L^{ab} M_{ab}+L^{\mu}_z Z_\mu+ L_z Z, \ee
where \bea &L^\mu=dx^\mu, ~~~ L^a=dx^a+dx^\mu {v}^{.
a}_{\mu},&
\nn\\
&L^{\mu \nu}=0,~~~ L^{\mu a}= dv^{\mu a},~~~ L^{ab}=0.&
\label{W1} \eea \bea &L^{\mu}_z= dc^{\mu} +
dx^{\nu}{\epsilon_{\nu}}^{\mu}-v^{\mu}_{. a}dx^a -\frac12
dx^\nu{v_{\nu a}}v^{\mu a},&
\nn\\
&L_z = dc + \frac12\epsilon_{\mu\nu}{v^{\mu}}_{. a}dv^{\nu
a} . \label{o-} \label{Lmunu}&\eea In the case of a four
dimensional space-time the extended Galilei algebra has a
further extension. In fact, owing to the existence of the
two dimensional Levi-Civita tensor, we can construct a
non-trivial closed invariant 2-form $\Omega_2^{\mu\nu}$,
that transforms as a symmetric tensor of order two under
the stability group \be
\Omega^{(\mu\nu)}_2=\epsilon_{ab}L^{(\mu a} L^{\nu)
b}=d(\epsilon_{ab}v^{\mu a} dv^{\nu b}).\ee The 1-form
$\epsilon_{ab} v^{\mu a} dv^{\nu b}$ is not invariant,
therefore the Eilenberg-Chevalley cohomology of degree two
is non trivial \cite{bargmann,levyleblond69,azcarragabook}.
This implies the possibility of further extending the
algebra. We denote the new generators of the algebra by
$\tilde{Z}_{(\mu\nu)}$, where $(\mu\nu)$ means symmetric
with respect the interchange of $\mu$ and $\nu$.

The new non vanishing commutation relations are \be \left[
M_{\mu a}, ~ M_{\nu b}\right] =
i\eta_{ab}\;\epsilon_{\mu\nu} Z+i
\epsilon_{ab}\tilde{Z}_{(\mu\nu)},\ee \be \left[
\tilde{Z}_{(\mu\nu)}, M_{\rho\s} \right] =
i\eta_{\nu[\rho}\tilde{Z}_{(\mu\s)]}
+i\eta_{\mu[\rho}\tilde{Z}_{(\nu\s)]}.\ee

 In the new extended algebra the MC 1-form
has one more term, given by $\frac{1}{2}
\tilde{L}^{\mu\nu}_z\tilde{Z}_{(\mu\nu)}$, and the coset
element is \be \label{p3} g=e^{i x^{\mu}P_{\mu}}\;e^{i
x^{a}P_{a}}e^{iv^{\mu a}M_{\mu a}} \;e^{ic^{\mu}Z_{\mu}}
e^{ic\, Z}\;e^{i\frac 12\tilde {c}^{(\mu\nu)}\tilde{
 Z}_{(\mu\nu)}}, \ee
where $\tilde {c}^{(\mu\nu)}$ are the group parameters
associated to generators $\tilde{Z}_{\mu\nu}$. The new term
in the MC 1-form is given by \be \tilde
L^{(\mu\nu)}=d\tilde c^{\mu\nu}+\frac 12\epsilon_{ab}
v^{a(\mu} dv^{\nu) b}.\ee

\section{Exotic String Action}

The ordinary non-relativistic string action can be written
as the pullback on the world-sheet with coordinates $(\tau,
\sigma)$ of the invariant 2-form \cite{Brugues:2006yd}
\be\label{E.16}\Omega_2=\epsilon_{\mu\nu}L^\mu
L_z^{\nu}=\epsilon_{\mu\nu}dx^\mu\left(dc^\nu-v^{\nu b}
dx^b+\frac 1 4 dx^\nu v_{\sigma a}v^{\sigma a}\right).\ee

In four dimensions  we can also consider a tensor valued
2-form \footnote{We could have considered also a scalar
2-form $\tilde\Omega=L_z\eta_{\mu\nu}
\tilde{L}^{(\lambda\nu)}$, however under double dimensional
reduction it does not produce the action of an exotic
particle} \be \Omega_2^{\rho\nu\lambda}= L^\rho
\tilde{L}^{(\lambda\nu)}=
dx^\rho\left(d\tilde{c}^{(\lambda\nu)}+\frac 1
2\epsilon_{ab} v^{(\lambda a}dv^{\nu) b}\right),\ee which
allows us to construct an action for an exotic string. If
we eliminate the closed 2-form $dx^\rho\,
d\tilde{c}^{\mu\nu}$ and we add the antisymmetric part in
$\lambda, \nu$, which is a closed form, we get
\be\Omega_2^{\rho\nu\lambda}=dx^\rho v^{\lambda a}dv^{\nu
b}\epsilon_{ab}.\ee

Notice that \be \Omega_3^{\rho\nu\lambda}=d(
\Omega_2^{\rho\nu\lambda})=L^\rho L^{\lambda
a}L^{b\nu}\epsilon_{ab}\ee is a closed invariant 3-form.
Since $\Omega_2^{\rho\nu\lambda}$ is not an invariant
2-form of the unextended string Galilei algebra we conclude
that the Eilenberg-Chevalley cohomology of degree three is
not trivial.


The string action is obtained by taking a linear
combination of the pullback of the previous 2-forms on the
world sheet. T is the string tension and
$\theta_{\rho\nu\lambda}$ is a spurion tensor field that
generalizes the magnetic field appearing in the case of a
particle in a magnetic field in $2+1$ dimensions.

\be S=T\int{d^2\sigma} \det[e]\left(v^{\mu a}
e_\mu^j\de_jx^a +\frac 12 v_{\mu a} v^{\mu
a}\right)+\theta_{\rho\nu\lambda}\int{d^2\sigma}\det[e]\epsilon^{\rho\mu}
\left(e_\mu^j v^{\lambda a}\de_j v^{\nu b}
\epsilon^{ab}\right)\label{eq:20}.\ee Here
$\de_i=(\de/\de\tau,\de/\de\sigma)$. The metric tensor is
 as before $\eta^{\mu\nu}=(-1,+1)$. The $e_i^\mu=\de_i x^\mu$ are the
zweibein, $e_\mu^j$ are the inverse zweibein, and
$\det[e]=\det[e_i^\mu]$ is the corresponding determinant.
Notice that $[T]=\ell^{-2}$ and $[\theta]=\ell^{-1}$. In
the previous equation $\theta_{\mu\rho\lambda}$ is a
spurion tensor.

If we want to get by double dimensional reduction the
action for the exotic particle
\cite{LSZ,DH1,JackN,DH2,HP1,PHMP,HP3,mariano1,OP,Horv} we
have to make a particular choice for the parameters
$\theta_{\rho\nu\lambda}$. By performing the dimensional
reduction of the exotic term in eq. (\ref{eq:20}) we find
\be \theta_{\rho\nu\lambda}\int d\tau\epsilon^{\rho 0}
v^{\lambda a} \de_\tau v^{\nu b}\epsilon_{ab},\ee to be
compared with the exotic term of the action for the
particle \be \kappa\int d\tau v^{0a}\de_\tau
v^{0b}\epsilon_{ab}.\ee Therefore the unique choice is to
take \be
\theta_{\rho\nu\lambda}=\theta_\lambda\epsilon_{\rho\nu}
\label{eq:23},\ee with $\theta_1=0$. Other possible forms
of the spurion field will not be discussed in this paper.

\section{Dynamics of the exotic string}

In this section we will study the dynamics of the exotic
string that under dimensional reduction produces the exotic
particle. We will use eq. (\ref{eq:23}) without the
restriction on $\theta_\mu$. In the static gauge, where we
make the identification $(x^0,x^1)=(\tau,\sigma)$, the
complete string action becomes

\be S=T\int{dx^0 dx^1} \left(v_{\mu a} \de^\mu x^a +\frac
12 v_{\mu a} v^{\mu a}\right)+\theta^\lambda\int{dx^0
dx^1}\left(\eta^{\mu\nu}\de_\nu v_{\mu b} v_{\lambda a}
\epsilon^{ab}\right)\label{action2}.\ee

The Lagrangian equations of motion are \be T\left(-\de_\mu
x_a +v_{\mu a}\right) +\theta_\mu \de^\sigma v_{\sigma b}
\epsilon^{ab}-\theta^\lambda\de_\mu
v_{\lambda_b}\epsilon^{ba}=0,\ee \be \de^\mu v_{\mu
a}=0.\ee

The boundary conditions in general imply \be
\label{boundary}(-T v_{1a}\delta x^a+\theta^\lambda
v_{\lambda
 a}\epsilon^{ab} \delta v_{1b})|^{L}_0 =0,\ee
where $L$ is the lenght of the string.

These conditions can be satisfied, for example, in the
following ways
\begin{enumerate}
\item By choosing both $x^a$ and $v^{\lambda a}$ as
periodic functions. \item By choosing $x^a$ periodic and
$v^{\lambda a}$ with fixed values at the boundaries. \item
By choosing periodic $v^{\lambda a}$ and  $x^a$ with fixed
values at the boundaries. \item By choosing both  $x^a$ and
$v^{\lambda a}$ with fixed values at the boundaries. \item
By choosing $v^{\lambda a}=0$ at the boundaries.
\end{enumerate}

Notice that the cases 1) and 5) correspond to a closed and
an open string respectively (case $\theta_\mu=0$). Notice
too that, for $\theta_{\mu} = 0$, one can eliminate the
variables $v$, which are essentially the derivative of the
variable $x$.

In order to understand which are the physical degrees of
freedom of the model and their dynamics we will now study
the hamiltonian formalism.

Let us start by  evaluating the canonical momenta: \be
\Pi_{\mu a}=\frac{\de {\cal L}}{\de \dot{v}^{\mu
a}},~~~p_a=\frac{\de{\cal L}}{\de \dot x^a},\ee One finds
the following primary constraints:
\bea\phi_a&=&\Pi_{0a}+\theta^\mu v_{\mu
b}\epsilon^{ab}=0,\nn\\
\chi_a&=&\Pi_{1a}=0,\nn\\
\xi_a&=&\frac 1{\sqrt{T}}\left (p_a + Tv_{0a}\right)=0,\eea
where $\Pi_{\mu a}$ are the momenta conjugated to $v_{\mu
a}$ and $P_a$ the ones conjugated to $x_a$, where, for a
generic field $\phi$,  $\dot\phi={\de\phi}/{\de t}=
{\de\phi}{\de x^0}$, with $x^0 = t$ .

The canonical Hamiltonian density turns out to be: \be
{\cal H}_c=p_a \dot x_a +\Pi_{\mu a}\dot v^{\mu a}-{\cal
L}= \frac 12
T\left(v_{0a}^2-v_{1a}^2\right)-Tv_{1a}x_a^\prime-\theta^0v_{0a}
v_{1b}^\prime \epsilon_{ab}-\theta^1
v_{1a}v_{1b}^\prime\epsilon^{ab},\label{eq:31}\ee where a
prime denotes derivative with respect to $\sigma$ (or
$x_1$).

It is easily verified that the following combinations of
primary constraints \be K_a=\frac
{(-1)^a}{\sqrt{T}}\left(+\theta_1\epsilon_{ab}\xi_b+\sqrt{T}\chi_a\right)\ee
are first class. By taking the independent combinations\be
S_a= \frac{(-1)^a}
z\left(\sqrt{T}\epsilon_{ab}\xi_b-\theta_1\chi_a\right),
~~~ z^2=T+\theta_1^2,\ee one sees that the primary
constraints separate as the first class, $K_a$, and the
second class ones $(\phi_a,S_a)$.

The matrix of the Poisson brackets among the second class
constraints, $(\phi_a,S_a)$, is given by: \be
\begin{array}{c|cccc}
& \phi_2 & \phi_3 & S_2 &S_3  \\
\hline
\phi_2 & 0 & -2\theta^0 & 0 & z \\
\phi_3 & 2\theta^0 & 0 & z & 0 \\
S_2 & 0 & -z & 0 & 0 \\
S_3 & -z & 0 & 0 & 0 \\
\end{array} \times \delta(\sigma-\sigma').
\ee \vskip0.2cm\noindent
The determinant of  the
$\sigma,\sigma^{\prime}$ independent part of this matrix
turns out to be proportional to $z^4$. We can now evaluate
the Dirac brackets with respect to these constraints.
Remember that the general form for the Dirac brackets,
given a set $(\psi_\alpha)$ of constraints, with a Poisson
bracket matrix given by $C_{\alpha,\beta}$, is \be
\{A,B\}^*=\{A,B\}-\sum_{\alpha,\beta}\{A,\psi_\alpha\}C^{-1}_{\alpha,\beta}
\{\psi_\beta,B\}\label{Dirac1}.\ee Just to give an example,
considering the set $(\phi_a,S_a)$, we find the  following
result \bea
\{x_a(\sigma,\tau),x_b(\sigma^\prime,\tau)\}^*&=&-\frac{2\theta^0}{z^4}
\epsilon_{ab}\delta(\sigma-\sigma'),\nn\\
\{v_{\mu a}(\sigma,\tau),v_{\nu
b}(\sigma^\prime,\tau)\}^*&=&-\frac{\theta_1}
{z^2}\epsilon_{ab}\left[\left(\eta_{\mu 0}\eta_{\nu
1}+\eta_{\mu 1}\eta_{\nu
0}\right)+\frac{2\theta^0\theta^1}{z^2}\eta_{\mu
1}\eta_{\nu 1}\right]\delta(\sigma-\sigma'),\nn\\
\{x_a(\sigma,\tau), v_{\mu
b}(\sigma^\prime,\tau)\}^*&=&\frac 1
{z^2}\delta_{ab}\left[\eta_{\mu 0}+
\frac{2\theta^0\theta^1}{z^2}\eta_{\mu
1}\right]\delta(\sigma-\sigma').\eea

 By requiring the stability of the primary
constraints, that is by requiring that the Poisson brackets
of the primary constraints with the hamiltonian be zero,
one finds the following secondary constraints:
\be\lambda_a=T\left(v_{1
b}+x_b^\prime\right)\epsilon_{ba}+\theta^\mu v_{\mu
a}^\prime=0.\label{eq:37}\ee With these two constraints the
primary first-class constraints $(K_a,\lambda_a)$ become
second-class. In fact, the Dirac matrix of the constrains
\be
\begin{array}{c|cccc}
& K_2 & K_3 & \lambda_2 &\lambda_3  \\
\hline
K_2 & 0 & 0 & 0 & -T \\
K_3 & 0 & 0 & -T & 0 \\
\lambda_2 & 0& T & 0 & -\frac{2T^2\theta^0\theta_1^2}{z^4} \\
\lambda_3 & T & 0 & \frac{2T^2\theta^0\theta_1^2}{z^4} & 0 \\
\end{array} \times \delta(\sigma-\sigma').
\label{matrix-secondary}\ee \vskip0.2cm\noindent has a
determinant proportional to $T^4\not=0$. We now iterate the
procedure to evaluate the final Dirac brackets, including
the new set of second class constraints. The final result
is (we write only the non vanishing Dirac brackets)\bea
\{x_a(\sigma,\tau),x_b(\sigma^\prime,\tau)\}^{**}&=&-\
\frac{2\theta^0}{T^2}\epsilon_{ab}\delta(\sigma-\sigma'),\nn\\
\{x_a(\sigma,\tau), v_{0
b}(\sigma^\prime,\tau)\}^{**}&=&-\frac
1T\delta_{ab}\delta(\sigma-\sigma'),\eea
\bea \{x_a(\sigma,\tau),p_b(\sigma^\prime,\tau)\}^{**}&=&\delta_{ab}\delta(\sigma-\sigma'),\nn\\
\{x_a(\sigma,\tau), \Pi_{0
b}(\sigma^\prime,\tau)\}^{**}&=&\frac{\theta^0}T\epsilon_{ab}\delta(\sigma-\sigma'),\eea
with the $^{**}$ brackets evaluated as in (\ref{Dirac1}),
using the inverse of the matrix (\ref{matrix-secondary}).
At this point we can use the full set of constraints and
the final phase space can be taken as $(x_a,p_a)$.

Notice that using the constraints $\lambda_a$ we can form
the combination \be v_{12}\lambda_3-\lambda_2 v_{13}=
Tv_{1a}^2+T v_{1 a} x_a^\prime -\theta^0v_{0a}^\prime
v_{1b}\epsilon^{ab}-\theta^1v_{1a}v_{1b}^\prime\epsilon^{ab}=0.\ee
By taking the canonical hamiltonian, see
equation(\ref{eq:31})\be H_c=\int d\sigma{\cal H}_c,\ee we
see that the last three terms of the previous identity
coincide with the last three terms in $H_c$, except for a
sign and a total derivative in $\sigma$, containing only
the $v^{\mu a}$ variables. This total derivative is zero
for periodic boundary conditions and it is a constant for
fixed boundary conditions. Therefore, apart from this
constant, we get \be H_c=\frac T2 \int d\sigma\left(
v_{0a}^2+v_{1a}^2\right).\label{eq:43}\ee

\section{Equivalence with the free non-relativistic string}

In order to express the hamiltonian in terms of the
independent variables $x^a$ and $p_a$ one has to solve the
constraints $\xi_a$ and $\lambda_a$. Since these
constraints in general depend on $v'_{1a}$, the hamiltonian
density that one gets is generally non-local. See appendix
A for the expression in terms of $x^a, p_a$. However, we
can show that this non-locality can be eliminated by a
convenient change of variables. Let us start considering
the simpler case $\theta_1=0$, where the constraints do not
depend on $v_{1a}'$. By solving the constraints we get \be
v_{12}=-x_2^\prime - \frac{\theta^0}{T^2}
p_3^\prime,~~~v_{13}=-x_3^\prime+\frac{\theta^0}{T^2}
p_2^\prime.\label{eq:21}\ee Therefore the hamiltonian is
given by \be H_c=\int d\sigma\left[\frac 1{2T} p_a^2+\frac
12 \frac{\theta^{0\, 2}}{T^3}p_a^{\prime\, 2}+\frac T2
x_a^{\prime\, 2}+\frac{\theta^0} T \epsilon^{ab}x_a^\prime
p_b^\prime\right].\ee
 From (\ref{eq:21}) we see that defining \be
y_2= x_2 +\frac{\theta^0}{T^2} p_3,~~~y_3=x_3
-\frac{\theta^0}{T^2} p_2,\ee the hamiltonian can be
written as \be H_c= \int d\sigma\left(
\frac{p_a^2}{2T}+\frac T2 y_a^{\prime\,2}\right).\ee Since
the variables $v_{1a}$ (and correspondingly $y_a$) have
zero Dirac brackets with themselves, the theory is
equivalent to a non relativistic free string
\cite{Gomis:2004ht}.

Coming back to the action in eq. (\ref{action2}) for the
general case $\theta_1\not=0$, if we integrate by parts the
last term (neglecting again the total derivative for the
reasons explained above)   we get \be S=T\int{dx^0 dx^1}
\left[v_{\mu a} \left(-\de^\mu
x^a-\frac{\theta^\lambda}T\de^\mu v_{\lambda b}
\epsilon^{ba}\right) +\frac 12 v_{\mu a} v^{\mu
a}\right],\ee By defining now  the new variables \be y^a=
x^a-\frac{\theta^\lambda}T v_{\lambda b} \epsilon^{ba},\ee
we see that the model is equivalent to the free non
relativistic string  \cite{Brugues:2004an} also for
$\theta^1\not=0$, since the new action coincides with the
first term in eq. (\ref{eq:20}). Also in this case the
Dirac brackets of the $y_a's$ vanish making the theory a
commmutative one.

However we have to consider the boundary conditions
discussed in Section IV. When $x_a$ and $y_a$ satisfy the
same boundary conditions, as for instance in  cases 1), 4)
and 5),  there is a complete equivalence of this exotic
string with the ordinary one. In the
 cases  2) and 3) the $y_a$ variables do not
satisfy definite boundary conditions and we loose the
equivalence. Notice also that, when the equivalence is
realized, the invariance group (in the $y_a$ variables) is
isomorphic to the extended Galilei group without the
extended generator $\tilde{Z}_{(\mu\nu)}$.

\section{Discussion}

In this paper we have constructed an exotic classical
non-relativistic string in 3+1 dimensions. The model
contains, apart from the ordinary non-relativistic action,
an extra term with a spurion field that exists only in 3+1
dimensions. The existence of this extra term is associated
to the existence of a new term in the extended Galilei
algebra in 4 dimensions. In general the model is
non-commutative in the sense that the Dirac bracket of the
physical coordinates are not vanishing. It is possible to
introduce a new set of coordinates in which the model
becomes commutative if the boundary conditions of the old
and the new coordinates coincide. We should point out that
the transformation properties of the old and the new
coordinates are different with respect to the Galilei
transformations.

This exotic string by double dimensional reduction
reproduces the exotic Galilei particle in 2+1 dimensions.

It should be noticed that the model considered in this
paper could be obtained from an exotic relativistic model
in 4 dimensions as non-relativistic limit.

We conclude by saying that it will be interesting to see if
exists an exotic string theory  for a non-flat case, in
such a way that under double dimensional  reduction it
could reproduce the exotic Newton-Hooke particle in 2+1
dimensions.

\appendix
\section{The Hamiltonian density in the reduced canonical space}

From the constraints $\lambda_a$ and $\xi_a$ the variables
$v_{\mu a}$ can be obtained as a function of the canonical
coordinates of the reduced space $\{x^a, p_a\}$. To this
end we must solve the following differential system
\be\label{2000} v_{12}^{\prime}(\sigma) = \omega
v_{13}(\sigma) + f_1(\sigma),~~~
 v_{13}^{\prime}(\sigma) = - \omega v_{12}(\sigma) + f_2(\sigma),
\ee \noindent where \be\label{2001} f_1 = \omega
(x^{\prime}_3 - \frac{\theta_{0}}{T^2} p_2^{\prime}),~~~
f_2 =  -\omega (x^{\prime}_2 + \frac{\theta_{0}}{T^2}
p_3^{\prime}). \ee The other variables $v_{0a}$ are easily
obtained from the constraints $\xi_a$. In these equations
we have put $\omega = T/\theta^1$. Observe that $\omega$
becomes infinite when $\theta^1$ is zero. In this limit the
system degenerates to an algebraic system, which was
already discussed in Section V. The system (\ref{2000}) can
be easily integrated and the solution is \be\label{2002}
v_{12} = \frac{1}{\sqrt{2}}(w_1 + w_2),~~~
v_{13}(\sigma) = \frac{i}{\sqrt{2}}(w_1 - w_2),\\
\nonumber \ee \noindent where \be\label{2003} w_1(\sigma) =
\frac{1}{\sqrt{2}}e^{i\omega\sigma}
\left(\int_0^{\sigma}e^{-i\omega\sigma^{\prime}}(f_1(\sigma^{\prime})
- if_2(\sigma^{\prime}))d\sigma^{\prime} + d\right),
~~~w_2(\sigma) = w_1(\sigma)^*. \ee In these equations the
complex constant of integration $d$ is determined by the
boundary conditions discussed in Section IV. With these
expressions one can evaluate the canonical Hamiltonian ${
H}_c$  (see eq. (\ref{eq:43})). The resulting hamiltonian
is manifestly non-local in the variables $(x^a,p_a)$.

\section*{Acknowledgements}

We are grateful to Jaume Gomis, Kiyoshi Kamimura for interesting
discussions. This work has been supported by the European EC-RTN
project MRTN-CT-2004-005104, MCYT FPA 2004-04582-C02-01 and CIRIT GC
2005SGR-00564. One of us, J.G. would like to thank the Galileo
Galilei Institute for Theoretical Physics for its hospitality and
INFN for partial support during part of the elaboration of this
work.


\begin{thebibliography}{99}

\bibitem{LL}
J.-M.~L\'evy-Leblond, {\it Galilei group and Galilean
invariance}. In: {\it Group Theory and Applications} (Loebl
Ed.), {\bf II}, Acad. Press, New York, p. 222 (1972).

\bibitem{BGO}
 A. Ballesteros, M. Gadella and
M.~del Olmo, {\it Moyal quantization of 2+1 dimensional
Galilean systems}. {\sl Journ. Math. Phys.} {\bf 33} (1992)
3379;

\bibitem{BGGK}
Y.~Brihaye, C.~Gonera, S.~Giller and P.~Kosi\'nski, {\it
Galilean invariance in $2+1$ dimensions.}
arXiv:hep-th/9503046.

\bibitem{bargmann}
 V.~Bargmann,
 \emph{``On Unitary Ray Representations Of Continuous Groups,''}
  Annals Math.\  {\bf 59} (1954) 1.


\bibitem{levyleblond69}
J.-M. L\'evy-Leblond, \emph{``Group-Theoretical Foundations
of Classical Mechanics: The Lagrangian Gauge Problem,"}
Comm. Math. Phys. {\bf 12} (1969) 64.


\bibitem{azcarragabook}
  J.~A.~de Azcarraga and J.~M.~Izquierdo,
\emph{``Lie groups, Lie algebras, cohomology and some
applications in physics.''} Cambridge Univ. Press, 1995.



\bibitem{LSZ}
  J.~Lukierski, P.~C.~Stichel and W.~J.~Zakrzewski,
  \emph{``Galilean-invariant (2+1)-dimensional models with a Chern-Simons-like  term
  and D = 2 noncommutative geometry,''}
  Annals Phys.\  {\bf 260} (1997) 224
  [arXiv:hep-th/9612017].


\bibitem{DH1}
  C.~Duval and P.~A.~Horvathy,
 \emph{ ``The "Peierls substitution" and the exotic Galilei group,''}
  Phys.\ Lett.\ B {\bf 479} (2000) 284
  [arXiv:hep-th/0002233];
  \emph{``Exotic Galilean symmetry in the non-commutative plane, and the Hall
  effect,''}
  J.\ Phys.\ A {\bf 34} (2001) 10097
  [arXiv:hep-th/0106089].


\bibitem{JackN}
  R.~Jackiw and V.~P.~Nair,
\emph{  ``Anyon spin and the exotic central extension of
the planar Galilei  group,''}
  Phys.\ Lett.\ B {\bf 480} (2000) 237
  [arXiv:hep-th/0003130].

\bibitem{DH2}
  C.~Duval, Z.~Horvath and P.~A.~Horvathy,
\emph{``Exotic plasma as classical Hall liquid,''}
  Int.\ J.\ Mod.\ Phys.\ B {\bf 15} (2001) 3397
  [arXiv:cond-mat/0101449].


\bibitem{HP1}
  P.~A.~Horvathy and M.~S.~Plyushchay,
\emph{``Non-relativistic anyons,  exotic Galilean symmetry
and noncommutative plane,''}
  JHEP {\bf 0206} (2002) 033
  [arXiv:hep-th/0201228].


\bibitem{PHMP}
 P.~A.~Horvathy and M.~S.~Plyushchay,
  \emph{``Anyon wave equations and the noncommutative plane,''}
  Phys.\ Lett.\ B {\bf 595} (2004) 547
  [arXiv:hep-th/0404137].



\bibitem{HP3}
  P.~A.~Horvathy and M.~S.~Plyushchay,
\emph{``Nonrelativistic anyons in external electromagnetic
field,''}
  Nucl.\ Phys.\ B {\bf 714} (2005) 269
  [arXiv:hep-th/0502040].

\bibitem{mariano1}
J. Negro, M. A. del Olmo and J. Tosiek, \emph{``Anyons,
group theory and planar physics,}" J. Math. Phys.
\textbf{47} (2006) 033508 [arXiv:math-ph/0512007].


\bibitem{OP}
 M.~A.~del Olmo and M.~S.~Plyushchay,
  \emph{``Electric Chern-Simons Term, Enlarged Exotic Galilei Symmetry And
  Noncommutative Plane,''}
 Annals Phys.\  {\bf 321} (2006) 2830
  [arXiv:hep-th/0508020].


\bibitem{Horv}
  P.~A.~Horvathy,
\emph{  ``Non-commutative mechanics, in mathematical and in
condensed matter physics,''}
  arXiv:cond-mat/0609571.

\bibitem{Polychronakos:2007df}
  A.~P.~Polychronakos,
  ``Noncommutative Fluids,''
  arXiv:0706.1095 [hep-th].

\bibitem{Mariano}
O. Arratia, M.~A.~Martin and M.~A.~Olmo, \emph{``Classical
Systems and Representation of (2+1) Newton-Hooke
Symmetries,''} arXiv:math-ph/9903013.

\bibitem{Alvarez:2007fw}
  P.~D.~Alvarez, J.~Gomis, K.~Kamimura and M.~S.~Plyushchay,
  ``(2+1)D exotic Newton-Hooke symmetry, duality and projective phase,''
  arXiv:hep-th/0702014.


\bibitem{Gomis:2000bd}
J.~Gomis and H.~Ooguri, ``Non-relativistic closed string
theory,'' J.\ Math.\ Phys.\ {\bf 42} (2001) 3127
[arXiv:hep-th/0009181].



\bibitem{Danielsson:2000gi}
U.~H.~Danielsson,  A.~Guijosa and   M.~Kruczenski, ``IIA/B,
wound and wrapped,'' JHEP {\bf 0010} (2000) 020
[arXiv:hep-th/0009182].

\bibitem{Coleman}
  S.~R.~Coleman, J.~Wess and B.~Zumino,
  \emph{``Structure of phenomenological Lagrangians. 1,''}
  Phys.\ Rev.\  {\bf 177} (1969) 2239;
  C.~G.~.~Callan, S.~R.~Coleman, J.~Wess and B.~Zumino,
  \emph{``Structure of phenomenological Lagrangians. 2,''}
  Phys.\ Rev.\  {\bf 177} (1969) 2247.





\bibitem{Brugues:2004an}
  J.~Brugues,   T.~Curtright,  J.~Gomis   and           L.~Mezincescu,
  ``Non-relativistic strings and branes as non-linear realizations of
  Galilei 
  [arXiv:hep-th/0404175].  








\bibitem{Brugues:2006yd}
  J.~Brugues, J.~Gomis and K.~Kamimura,
  ``Newton-Hooke algebras, non-relativistic branes and generalized pp-wave
  Phys.\ Rev.\  D {\bf 73} (2006) 085011
  [arXiv:hep-th/0603023].

\bibitem{GomisKW}
  J.~Gomis, K.~Kamimura and P.~West,
  \emph{``The construction of brane and superbrane actions using non-linear
  realisations,''}
  Class.\ Quant.\ Grav.\  {\bf 23} (2006) 7369
  [arXiv:hep-th/0607057].

\bibitem{Gauntlett:1990nk}
  J.~P.~Gauntlett, J.~Gomis and P.~K.~Townsend,
  ``Particle Actions As Wess-Zumino Terms For Space-Time (Super)Symmetry
  Groups,''
  Phys.\ Lett.\  B {\bf 249} (1990) 255.



\bibitem{Gomis:2004ht}
  J.~Gomis and F.~Passerini,
  ``Rotating solutions of non-relativistic string theory,''
  Phys.\ Lett.\  B {\bf 617} (2005) 182
  [arXiv:hep-th/0411195].

\end{thebibliography}
\end{document}